\def\tr{{\text{tr}}\,}

\def\epsilonF{\epsilon_{\text{F}}}
\def\kF{k_{\text{F}}}
\def\vF{v_{\text{F}}}
\def\NF{N_{\text{F}}}
\def\me{m_{\text{e}}}

\def\sgn{{\text{sgn\,}}}

\def\be{\begin{equation}}
\def\ee{\end{equation}}
\def\bea{\begin{eqnarray}}
\def\eea{\end{eqnarray}}
\def\bse{\begin{subequations}}
\def\ese{\end{subequations}}

\documentclass[prl,twocolumn,showpacs,preprintnumbers,amsmath,amssymb]{revtex4}
\usepackage{graphicx}
\usepackage{dcolumn}
\usepackage{bm}
\begin{document}
\preprint{}
\title{Theory of a Fermi-Liquid-to-Non-Fermi-Liquid Quantum Phase Transition in Dimensions $d>1$}
\author{T.R. Kirkpatrick$^{1}$ and D. Belitz$^{2}$}
\affiliation{$^{1}$Institute for Physical Science and Technology and Department
                   of Physics, University of Maryland, College Park, MD 20742\\
         $^{2}$Department of Physics and Theoretical Science Institute, University
                of Oregon, Eugene, OR 97403}
\date{\today}
\begin{abstract}
We develop a theory for a generic instability of a Fermi liquid in dimension $d>1$ against the formation of a Luttinger-liquid-like state. The density of states at the Fermi level is the order parameter for the ensuing quantum phase transition, which is driven by the effective interaction strength. A scaling theory in conjunction with an effective field theoy for clean electrons is used to obtain the critical behavior of observables. In the Fermi-liquid phase the order-parameter susceptibility, which is measurable by tunneling, is predicted to diverge for $1<d<3$.

\end{abstract}
\maketitle
Landau's Fermi-liquid theory provides a very successful paradigm in condensed matter physics. By mapping the low-lying excitations in interacting Fermi systems onto those of noninteracting ones \cite{Baym_Pethick_1991} it explains many properties of electrons in solids, including the linear temperature ($T$) dependence of the specific heat, and the quadratic $T$-dependence of the electrical resistivity \cite{AGD_1963}. In a renormalization-group (RG) context it can be understood as the scaling behavior near a stable fixed point (FP) that governs the low-$T$ behavior of the system \cite{Shankar_1994}. Because of this success, deviations from Fermi-liquid (FL) behavior have attracted considerable attention \cite{ITP_Conference}. Examples include parts of the normal phase of high-$T_{\text{c}}$ superconductors \cite{Lee_Nagaosa_Wen_2006}, heavy-fermion systems \cite{Gegenwart_Si_Steglich_2008}, and the paramagnetic phase of the helimagnet MnSi at low $T$ \cite{Pfleiderer_Julian_Lonzarich_2001}.

There are different sources for non-Fermi-liquid (NFL) behavior. One is the vicinity of a quantum critical point where the low-$T$ behavior is governed by a critical FP rather than the stable FL FP, which changes the scaling behavior. The electrons couple to the critical soft modes, and the NFL behavior is confined to a small region in parameter space. This is believed to be the source of the observed NFL behavior in heavy-fermion systems \cite{Gegenwart_Si_Steglich_2008}. Another possibility is the existence of Goldstone modes due to a spontaneously broken symmetry and resulting long-range order. If the electrons couple to the Goldstone modes, NFL behavior can result in an entire phase. This has been proposed to explain the behavior of MnSi \cite{Kirkpatrick_Belitz_2010}. More generic mechanisms for NFL behavior, that do not rely on underlying long-range order, are hard to find. It is well known that in one-dimensional ($1$-$d$) fermion systems an arbitrarily small repulsive interaction amplitude $K_{\text{s}}$ (we will restrict ourselves to a point-like interaction in the spin-singlet particle-hole channel) leads to an instability of the FL against a Luttinger liquid (LL) that has a vanishing density of states (DOS) at the Fermi level \cite{Giamarchi_2004}. The LL is characterized by sound-like excitations, and its properties, apart from the vanishing DOS, are very similar to those of a FL. 
A natural question is whether in dimensions $d>1$ a similar instability will occur for $K_{\text{s}}$ greater than some $K_{\text{s}}^{\text c} > 0$. Despite substantial efforts, to date no description of such an instability has been found.

There are, however, indications that an instability exists. Perturbation theory in the FL phase yields nonanalytic dependencies on $T$, or the wave number $k$, for, e.g., the spin susceptibility and the specific heat coefficient \cite{Belitz_Kirkpatrick_Vojta_1997, Chitov_Millis_2001,  Chubukov_Maslov_2003} . For generic $d$ they take the form $T^{d-1}$ or $k^{d-1}$, with multiplicative logarithms in odd $d$. For $d=1$, the logarithmic divergencies coincide with the well-known perturbative signatures of the 
LL \cite{Solyom_1979}. This is reminiscent of disordered electrons, where perturbation theory generically yields a $T^{(d-2)/2}$ or $k^{d-2}$ behavior, which turns into $\log T$ or $ \log k$ in $d=2$. These perturbative ``weak-localization'' effects signalize the instability of the disordered FL against an Anderson or Anderson-Mott insulator \cite{Lee_Ramakrishnan_1985, Belitz_Kirkpatrick_1994}. In $d=2$ this instability occurs at arbitrarily small values of the disorder, whereas in $d>2$ a metal-insulator transition occurs at a nonzero critical value of the disorder. It is thus natural to speculate that a transition from a FL to a LL can occur in $d>1$.

In this Letter we construct a theory that describes a quantum phase transition from a FL to a NFL state with a vanishing DOS at the Fermi level in $d>1$. The DOS serves as the order parameter (OP) for the transition; the FL is the ordered phase. $d_{\text c}^- = 1$ is the lower critical dimension for the transition; fluctuations destroy the ordered phase for $d \leq d_{\text c}^-$. For $d=1+\epsilon$ ($\epsilon \ll 1$) the theory is controlled and the critical value of the interaction strength is $K_{\text{s}}^{\text c} = O(\epsilon^{1/2})$. For larger $d$ the critical behavior is obtained from scaling considerations. Our approach is inspired by the nonlinear sigma-model description of the classical magnetic Heisenberg transition near $d=2$ \cite{Zinn-Justin_1996}. In many respects our theory is analogous to that of the Anderson-Mott metal-insulator transition of disordered interacting electrons, for which $d_{\text c}^- = 2$ as well \cite{Finkelstein_1983, Belitz_Kirkpatrick_1994},
even though the transition in disordered systems is to a non-standard insulator, while in clean ones, it is to a non-standard metal \cite{DOS_footnote}.

To identify the DOS as the OP for the FL-to-NFL transition we consider a Ward identity that reflects the broken symmetry between retarded and advanced degrees of freedom in a FL. It relates a two-particle correlation function $F_2$ (schematically, $\langle {\bar\psi}{\bar\psi}\psi\psi\rangle$, with $\bar\psi$ and $\psi$ fermion fields; a more explicit expression of $F_2$ will be given in Eq.\ (\ref{eq:12})) to a single-particle function $F_1$ ($\langle{\bar\psi}\psi\rangle$) and is a generalization of a Ward identity first considered for noninteracting electrons with quenched disorder \cite{Maleyev_Toperverg_1975, Schaefer_Wegner_1980}. With ${\bm p}$ the center-of-mass wave vector ($\vert{\bm p}\vert \approx \kF$; we denote the Fermi wave number, energy, and velocity by $\kF$, $\epsilonF$, and $\vF$), ${\bm k}$ the hydrodynamic wave vector ($\vert{\bm k}\vert \equiv k \ll \kF$), and $\me$ the electron mass it can be written
\bse
\label{eqs:1}
\bea
&&\hskip -110pt \left(i\Omega_{n_1-n_2} + {\bm p}\cdot{\bm k}/\me\right) F_2\left({\bm p},{\bm k};i\omega_{n_1},i\omega_{n_2}\right) 
\nonumber\\
\hskip 50pt&=& F_1\left({\bm p},{\bm k};i\omega_{n_1},i\omega_{n_2}\right),
\label{eq:1a}
\eea
$F_1$ is proportional to the difference between Green functions taken at the fermionic Matsubara frequencies $i\omega_{n_1}$ and $i\omega_{n_2}$. For $i\Omega_{n_1-n_2} = i\omega_{n_1} - i\omega_{n_2} \to 0$, ${\bm k} \to 0$, $F_1$ vanishes if 
$\omega_{n_1}\omega_{n_2} > 0$, but is nonzero if 
$\omega_{n_1}\omega_{n_2} < 0$. In the latter case, and for noninteracting electrons,
\be
F_1\left({\bm p},{\bm k};i\omega_{n_1},i\omega_{n_2}\right) \propto i\,\sgn(\Omega_{n_1-n_2})\,\delta({\epsilon_{\bm p}}-\epsilonF),
\label{eq:1b}
\ee
\ese
with $\epsilon_{\bm p}$ the single-particle energy-momentum relation. For $\omega_{n_1}\omega_{n_2} < 0$ there thus is a family of 4-fermion functions that diverge in the hydrodynamic limit of vanishing $\Omega_{n_1-n_2}$ and ${\bm k}$. Taking moments with respect to ${\bm p}$ yields an infinite number of soft modes, provided the density of states at the Fermi level, $N_{\text F} \propto \sum_{\bm p} \delta(\epsilon_{\bm p}-\epsilonF)$, is nonzero. In the presence of quenched disorder, in contrast, only the zeroth moment of Eq.\ (\ref{eq:1a}) 
is soft. These soft modes are the Goldstone modes of a spontaneously broken rotational symmetry in frequency space that is a nonlocal generalization of the one considered in Ref.\ \onlinecite{Belitz_Kirkpatrick_1997}. 

Equations.\ (\ref{eqs:1}) can be generalized to interacting systems along the lines of Ref.\ \cite{Belitz_Kirkpatrick_1997}. 
FL theory ensures that the basic structure of the identity is unchanged: The symmetry is broken, and Goldstone modes exist, as long as the DOS at the Fermi level is nonzero. The noninteracting DOS, $N_{\text F}$, gets replaced by the physical DOS, $N(\epsilonF)$, and the prefactor of the frequency $\Omega_{n_1-n_2}$ acquires a FL correction. $F_2$ remains massless, and the frequency continues to scale as a wave number. Conversely, a vanishing DOS implies that the symmetry is restored and the Goldstone modes have zero weight. If this happens, by varying some control parameter, with the Fermi level inside the conduction band, then the system will undergo a symmetry-restoring phase transition from a FL (ordered phase) to a NFL (disordered phase) with the DOS as the OP. In the FL the Goldstone modes are all proportional to the basic Goldstone propagator
\bse
\label{eqs:2}
\be
{\cal D}({\bm k},i\Omega) = N(\epsilonF)\,\varphi(i\Omega/\vF k)/k.
\label{eq:2a}
\ee
The explicit form of the function $\varphi$ depends on the dimensionality. In the limiting case $d \to 1$ one has
\be
\varphi(x) \propto \vert x\vert/(1+x^2).
\label{eq:2b}
\ee
\ese

These considerations show that in the ordered phase there are soft modes whose frequency scales as a wave number, $\Omega \sim k$: A dynamical exponent $z=1$ is associated with the stable FL FP. At a symmetry-restoring transition, described by a critical FP, $z \neq 1$ in general. 

The Goldstone modes are {\em not} related to the density propagator, which is governed by particle-number conservation. The latter, plus the fact that the thermodynamic density susceptibility $\partial n/\partial\mu$ is expected to show no critical behavior (see below), implies that in the density propagator one has $\Omega \sim k$, or $z=1$, at both the FL and the critical FPs. This is consistent with the fact that $\Omega \sim k$ at the stable FP that describes a LL in $d=1$ \cite{Giamarchi_2004}. Hence there is more than one dynamical exponent: The critical dynamical exponent $z$ ($z \neq 1$ in general), related to the Goldstone modes, and a second dynamical exponent $z_{\text{c}} = 1$ related to the charge or density dynamics. 

In what follows, we construct a scaling theory for a symmetry-restoring FL-to-NFL quantum phase transition where the DOS vanishes. We have also derived an effective field theory that allows for an explicit description of such a transition, the most important aspects of which we will sketch at the end of this Letter. 

We start by considering the free energy density $f$, which quite generally satisfies a scaling relation
\be
f(t,T,h) = b^{-(d+z)}\,f(t\,b^{1/\nu}, T\,b^z, h\,b^{y_h})\ .
\label{eq:3}
\ee
We have assigned scale dimensions $[L] = -1$ and $[T] = z$ to factors of length and temperature, energy, or inverse time ($\hbar = k_{\text B} = 1$), which yields $[f] = -d-z$ for the scale dimension of $f$ \cite{multiple_z_footnote}. $b$ is the RG length rescaling factor.  $h$ is the field conjugate to the OP, with scale dimension $[h] = y_h$. $t$ is the dimensional distance from the critical point, and $\nu = 1/[t]$ is the correlation length exponent. For the OP density $N = -(\partial f/\partial h)/T$ Eq.\ (\ref{eq:3}) implies
\bse
\label{eqs:4}
\bea
N(t,T) &=& b^{-d+y_h}\,N(t\,b^{1/\nu}, T\,b^z)\ .
\label{eq:4a}\\
&=& b^{-d + z}\,N(t\,b^{1/\nu}, T\,b^z)\ .
\label{eq:4b}
\eea
\ese 
Equation (\ref{eq:4b}) reflects the fact that $N$ is the DOS, which scales as an inverse energy times an inverse volume. Hence $y_h = z$. At $T=0$ and at criticality, respectively, the OP vanishes as a power law,
\bse
\be
N(t, T=0) \propto t^{\beta}\quad,\quad N(t=0,T) \propto T^{(d-z)/z}\ .
\label{eq:6a}
\ee
with a critical exponent
\be
\beta = \nu (d-z).
\label{eq:6b}
\ee
\ese
The corresponding results for the specific heat coefficient $\gamma$ are obtained from $C_V = \gamma\,T = -T \partial^2 f/\partial T^2$. The scaling behavior of $\gamma$ is the same as that of the DOS:
\be
\gamma(t,T) = b^{-d+z}\,\gamma(t\,b^{1/\nu}, T\,b^z),
\label{eq:7}
\ee
We next consider the OP susceptibility $\chi = \partial N/\partial h$ as a function of $t$, $T$, and the wave number $k$. In general,
\bse
\label{eqs:7'}
\be
\chi(t,T;k) = b^{2-\eta}\,\chi(t\,b^{1/\nu}, T\,b^z, kb),
\label{eq:7'a}
\ee
which defines the exponent $\eta$. At $T=0$ and at criticality, respectively, the homogeneous OP susceptibility diverges:
\bea
\chi(t,T=0,k=0) &\propto& t^{-\gamma}\quad,\quad\gamma = \nu (2 - \eta),
\nonumber\\
\chi(t=0,T,k=0) &\propto& T^{-(2-\eta)/z}.
\label{eq:7'b}
\eea
\ese
From Eqs.\ (\ref{eq:3}), (\ref{eqs:4}), and (\ref{eq:7'a}) we find the exponent relation
\be
z = (d - \eta + 2)/2.
\label{eq:7''}
\ee
This implies that there are only two independent critical exponents, e.g., $\nu$ and $z$ (see, however, the remark above regarding multiple exponents $z$) rather than three as is generally the case at a quantum critical point \cite{Sachdev_1999}. 
For $\partial n/\partial\mu$ we expect no critical behavior since it does not show the perturbative nonanalyticities that are precursors of the critical behavior of other observables \cite{Belitz_Kirkpatrick_Vojta_1997, Belitz_Kirkpatrick_Vojta_2002}.
The scaling behavior of the electrical conductivity $\sigma = D_{\text c}\,\partial n/\partial\mu$ is therefore given by that of the charge diffusion coefficient $D_{\text c}$, which scales as a length squared divided by a time. Since $D_{\text{c}}$ describes the charge or density dynamics the relevant dynamical exponent in this context is $z_{\text{c}} = 1$. We thus have \cite{z_c_footnote}
\bse
\label{eqs:8}
\be
\sigma(t,T) = b^{2-z_{\text{c}}}\,\sigma(t\,b^{1/\nu}, T\,b^{z},T\,b^{z_{\text{c}}}).
\label{eq:8a}
\ee
If $z<1$ (see below) this yields for the electrical resistivity $\rho = 1/\sigma$ at criticality
\be
\rho(t=0,T) \propto T.
\label{eq:8b}
\ee
\ese

The preceding scaling predictions all pertain to the critical FP. Also of interest are the OP and the OP susceptibility in the ordered phase, $\vert t\vert = \infty$, where $\eta = d$, which implies $z=1$. From Eq.\ (\ref{eq:4b}) we have
\be
N(\vert t\vert = \infty,T) \propto \text{const.} + T^{d-1}.
\label{eq:8'}
\ee
This is one example of the perturbative nonanalyticities mentioned above. The same power law holds at $T=0$ as a function of the distance $\omega$ from the Fermi surface: $N(T=0,\omega) \propto \text{const.} + \omega^{d-1}$.  It is analogous to the Coulomb anomaly in disordered systems, where $N(T=0,\omega) \propto \text{const.} + \omega^{(d-2)/2}$ \cite{Altshuler_Aronov_1979}. The latter is a precursor of the quantum phase transition in disordered systems (the Anderson-Mott transition \cite{Belitz_Kirkpatrick_1994}), where the DOS vanishes and serves as an OP \cite{Belitz_Kirkpatrick_1995}. The current theory suggests that an analogous statement holds in clean ones. For the OP susceptibility we find from Eq.\ (\ref{eq:7'a})
\be
\chi(\vert t\vert = \infty,T,k) = k^{d-2}\,f_{\chi}(T/k) \propto T/k^{3-d}.
\label{eq:10}
\ee
In the second relation we have used the result of an explicit calculation \cite{us_tbp}, which yields $f_{\chi}(x \to 0) \propto x$. This divergence of the OP susceptibility, or the 2-point local-DOS correlation function,  which is observable by tunneling experiments, is a consequence of the Goldstone modes. It is analogous to the $1/k^{4-d}$ divergence of the longitudinal susceptibility in the ordered phase of a Heisenberg ferromagnet \cite{Brezin_Wallace_1973}. For a $2$-$d$ FL it predicts a $1/k$ divergence with a prefactor that is linear in $T$.

The preceding scaling considerations are expected to be valid between the lower critical dimension $d_{\text{c}}^- = 1$ and some upper critical dimension $d_{\text{c}}^+$. Equation (\ref{eq:10}) suggests $d_{\text{c}}^+ = 3$, but this requires further corroboration. For $d > d_{\text{c}}^+$ one expects 
the critical behavior to be mean-field like and governed by a Gaussian FP. 

We now sketch the derivation of an effective field theory that allows for an explicit description of a quantum phase transition of the type we have discussed above. A complete account will be given elsewhere \cite{us_tbp}. This effective theory is in the spirit of the matrix field theories that were pioneered by Wegner \cite{Wegner_1979, Schaefer_Wegner_1980}, and generalized by others \cite{Finkelstein_1983, Belitz_Kirkpatrick_1997}, for disordered systems. We consider a fermionic action and define electron bispinors
\be
\eta_n({\bm x}) = \left({\bar\psi}_{n\uparrow}({\bm x}),{\bar\psi}_{n\downarrow}({\bm x}),\psi_{n\downarrow}({\bm x}),-\psi_{n\uparrow}({\bm x})\right)/\sqrt{2}
\label{eq:11}
\ee
where ${\bar\psi}$ and $\psi$ are fermionic fields with Matsubara frequency index $N$ and spin projection $\uparrow\downarrow$, as well as adjoints $\eta_n^+({\bm x}) = C\eta_n({\bm x})$ with $C = i\sigma_1\otimes\sigma_2$, where $\sigma_{1,2}$ are Pauli matrices. We confine the tensor product $\eta_n^+({\bm x})\otimes\eta_m({\bm y})$ to a spin-quaternion-valued bosonic matrix field $Q_{nm}({\bm x},{\bm y})$ by means of a Lagrange multiplier $\Lambda_{nm}({\bm x},{\bm y})$. The Ward identity then takes the form of Eqs.\ (\ref{eqs:1}) with
\bea
F_2({\bm p},{\bm k};i\omega_{n_1},i\omega_{n_2}) &=& \langle \tr Q_{n_2 n_1}({\bm p}+{\bm k}/2, {\bm p}-{\bm k}/2)\,
\nonumber\\
&&\hskip -30pt
\times\tr Q_{n_1 n_2}({\bm p}-{\bm k}/2, {\bm p}+{\bm k}/2)\rangle
\label{eq:12}
\eea
where $\omega_{n_1}\omega_{n_2} < 0$. This identifies $q_{nm}({\bm p}_1,{\bm p}_2) \equiv \Theta(-nm)\,Q_{nm}({\bm p}_1,{\bm p}_2)$ as the Goldstone modes. The corresponding elements $\lambda$ of the Lagrange multiplier matrix field $\Lambda$ are also soft modes. The electron-electron interaction couples $q$ to $P_{nm}({\bm p}_1,{\bm p}_2) \equiv \Theta(nm)\,Q_{nm}({\bm p}_1,{\bm p}_2)$, and integrating out $P$, and the corresponding part of $\Lambda$, generates terms to all orders in $q$ and $\lambda$. An analogous procedure in the presence of quenched disorder provides a perturbative derivation, order by order in powers of $q$, of the generalized nonlinear sigma-model for the Anderson-Mott transition problem \cite{Finkelstein_1983, Belitz_Kirkpatrick_1994}. 
We have derived the complete action to order $q^4$, which suffices for a 1-loop RG calculation. The effects of $\lambda$ can be absorbed into diagram rules. The same method can be used to systematically derive higher-order terms. 

This effective theory can be analyzed by RG methods in $d=1+\epsilon$ in analogy to the disordered case in $d=2+\epsilon$ \cite{Belitz_Kirkpatrick_1994}. The 1-point function is proportional to the DOS:
\be
P^{(1)} = \langle \tr Q_{nn}({\bm x},{\bm x})\rangle\bigr\vert_{i\omega_n \to i0}\ \propto N(\epsilonF) \equiv N_{\text{F}}\,Z^{1/2}.
\label{eq:13}
\ee
Physically, $Z^{1/2} = (1 + \delta Z)^{1/2}$ is the physical DOS normalized by the bare or free-electron DOS; technically, it is the field-renormalization constant. It is related to, but not the same as, the residue of the pole in the Green function.  At 1-loop order one finds that $\delta Z$ is negative, logarithmically divergent in $d=1$, and proportional to $1/\epsilon$ in $d=1+\epsilon$. As a function of $K_{\text{s}}$ it is of $O(K_{\text{s}}^2)$ for small $K_{\text{s}}$. In a naive extrapolation the DOS thus vanishes at a critical value $K_{\text{s}}^{\text{c}} = O(\epsilon^{1/2})$. The 2-point function
\bse
\label{eqs:14}
\be
P^{(2)} = \langle q_{n_1n_2}({\bm k}_1,{\bm k}_2)\,q_{n_3n_4}({\bm k}_3,{\bm k}_4) \rangle
\label{eq:14a}
\ee
has a contribution proportional to $\delta_{n_1 n_3}\,\delta_{n_2 n_4}$ that constitutes the Goldstone propagator ${\cal D}$, Eqs.\ (\ref{eqs:2}). In $d=1$
\be
{\cal D}({\bm k},i\Omega) = Z\,H\,\vert\Omega\vert/({\bm k}^2/G^2 + H^2\Omega^2).
\label{eq:14b}
\ee
\ese
The bare values of $G$ (which is the loop expansion parameter) and $H$ are $1/\vF\NF$ and $\NF$, respectively. An explicit calculation finds no singular renormalizations of $G$ and $H$ at 1-loop order. 
Structural considerations confirm this and show that at 2-loop order there is a singular renormalization of $G$ due to insertion diagrams; an inspection of skeleton diagrams will require a 2-loop calculation. This strongly suggest a critical FP at 2-loop order with a FP value of the renormalized coupling constant $g = b^{\epsilon} G$ given by $g^* = O(\epsilon^{1/2})$. Choosing the independent exponents to be $\nu$ and $z$ this leads to
\be
\nu = 1/2\epsilon + O(1) \quad,\quad z = 1 + O(\epsilon).
\label{eq:15}
\ee
The $O(\epsilon)$ term in $z$ requires a 2-loop calculation; an educated guess is as follows. In the bare theory, $G^2 H \propto \me/n$, with $n$ the electron density. This quantity one does not expect to be renormalized, so $H \sim G^{-2} \sim b^{-2\epsilon}$, or $z = 1 - \epsilon$ \cite{z_footnote}. Hence $z < z_{\text{c}}$, which implies Eq.\ (\ref{eq:8b}). 

We stress that the RG theory, although it relies on an $\epsilon$-expansion about $d=1$, is not tied to the special properties of $1$-$d$ fermion systems \cite{Giamarchi_2004}. It only requires Goldstone modes with $k \sim \Omega$, which exist in any $d>1$, and we can apply a RG scheme that works in a fixed dimension \cite{Parisi_1980} $d>1$.

This work was supported by the National Science Foundation under Grant Nos.
DMR-09-29966, and DMR-09-01907.


\begin{thebibliography}{32}
\expandafter\ifx\csname natexlab\endcsname\relax\def\natexlab#1{#1}\fi
\expandafter\ifx\csname bibnamefont\endcsname\relax
  \def\bibnamefont#1{#1}\fi
\expandafter\ifx\csname bibfnamefont\endcsname\relax
  \def\bibfnamefont#1{#1}\fi
\expandafter\ifx\csname citenamefont\endcsname\relax
  \def\citenamefont#1{#1}\fi
\expandafter\ifx\csname url\endcsname\relax
  \def\url#1{\texttt{#1}}\fi
\expandafter\ifx\csname urlprefix\endcsname\relax\def\urlprefix{URL }\fi
\providecommand{\bibinfo}[2]{#2}
\providecommand{\eprint}[2][]{\url{#2}}

\bibitem[{\citenamefont{Baym and Pethick}(1991)}]{Baym_Pethick_1991}
\bibinfo{author}{\bibfnamefont{G.}~\bibnamefont{Baym}} \bibnamefont{and}
  \bibinfo{author}{\bibfnamefont{C.}~\bibnamefont{Pethick}},
  \emph{\bibinfo{title}{Landau Fermi-Liquid Theory}}
  (\bibinfo{publisher}{Wiley, New York}, \bibinfo{year}{1991}).

\bibitem[{\citenamefont{Abrikosov et~al.}(1963)\citenamefont{Abrikosov, Gorkov,
  and Dzyaloshinski}}]{AGD_1963}
\bibinfo{author}{\bibfnamefont{A.~A.} \bibnamefont{Abrikosov}},
  \bibinfo{author}{\bibfnamefont{L.~P.} \bibnamefont{Gorkov}},
  \bibnamefont{and} \bibinfo{author}{\bibfnamefont{I.~E.}
  \bibnamefont{Dzyaloshinski}}, \emph{\bibinfo{title}{Methods of Quantum Field
  Theory in Statistical Physics}} (\bibinfo{publisher}{Dover (New York)},
  \bibinfo{year}{1963}).

\bibitem[{\citenamefont{Shankar}(1994)}]{Shankar_1994}
\bibinfo{author}{\bibfnamefont{R.}~\bibnamefont{Shankar}},
  \bibinfo{journal}{Rev. Mod. Phys.} \textbf{\bibinfo{volume}{66}},
  \bibinfo{pages}{129} (\bibinfo{year}{1994}).

\bibitem[{ITP()}]{ITP_Conference}
\bibinfo{note}{See, e.g., {\it Institute for Theoretical Physics Conference on
  Non--Fermi Liquid Behavior in Metals}, P. Coleman, B. Maple, and A. Millis
  (eds.), J. Phys. Cond. Matt. {\bf 48}, No. 8 (1996).}

\bibitem[{\citenamefont{Lee et~al.}(2006)\citenamefont{Lee, Nagaosa, and
  Wen}}]{Lee_Nagaosa_Wen_2006}
\bibinfo{author}{\bibfnamefont{P.~A.} \bibnamefont{Lee}},
  \bibinfo{author}{\bibfnamefont{N.}~\bibnamefont{Nagaosa}}, \bibnamefont{and}
  \bibinfo{author}{\bibfnamefont{X.-G.} \bibnamefont{Wen}},
  \bibinfo{journal}{Rev. Mod. Phys.} \textbf{\bibinfo{volume}{78}},
  \bibinfo{pages}{17} (\bibinfo{year}{2006}).

\bibitem[{\citenamefont{Gegenwart et~al.}(2008)\citenamefont{Gegenwart, Si, and
  Steglich}}]{Gegenwart_Si_Steglich_2008}
\bibinfo{author}{\bibfnamefont{P.}~\bibnamefont{Gegenwart}},
  \bibinfo{author}{\bibfnamefont{Q.}~\bibnamefont{Si}}, \bibnamefont{and}
  \bibinfo{author}{\bibfnamefont{F.}~\bibnamefont{Steglich}},
  \bibinfo{journal}{Nature Physics} \textbf{\bibinfo{volume}{4}},
  \bibinfo{pages}{186} (\bibinfo{year}{2008}).

\bibitem[{\citenamefont{Pfleiderer et~al.}(2001)\citenamefont{Pfleiderer,
  Julian, and Lonzarich}}]{Pfleiderer_Julian_Lonzarich_2001}
\bibinfo{author}{\bibfnamefont{C.}~\bibnamefont{Pfleiderer}},
  \bibinfo{author}{\bibfnamefont{S.~R.} \bibnamefont{Julian}},
  \bibnamefont{and} \bibinfo{author}{\bibfnamefont{G.~G.}
  \bibnamefont{Lonzarich}}, \bibinfo{journal}{Nature (London)}
  \textbf{\bibinfo{volume}{414}}, \bibinfo{pages}{427} (\bibinfo{year}{2001}).

\bibitem[{\citenamefont{Kirkpatrick and
  Belitz}(2010)}]{Kirkpatrick_Belitz_2010}
\bibinfo{author}{\bibfnamefont{T.~R.} \bibnamefont{Kirkpatrick}}
  \bibnamefont{and} \bibinfo{author}{\bibfnamefont{D.}~\bibnamefont{Belitz}},
  \bibinfo{journal}{Phys. Rev. Lett.} \textbf{\bibinfo{volume}{104}},
  \bibinfo{pages}{256404} (\bibinfo{year}{2010}).

\bibitem[{\citenamefont{Giamarchi}(2004)}]{Giamarchi_2004}
\bibinfo{author}{\bibfnamefont{T.}~\bibnamefont{Giamarchi}},
  \emph{\bibinfo{title}{Quantum Physics in One Dimension}}
  (\bibinfo{publisher}{Clarendon, Oxford}, \bibinfo{year}{2004}).

\bibitem[{\citenamefont{Belitz et~al.}(1997)\citenamefont{Belitz, Kirkpatrick,
  and Vojta}}]{Belitz_Kirkpatrick_Vojta_1997}
\bibinfo{author}{\bibfnamefont{D.}~\bibnamefont{Belitz}},
  \bibinfo{author}{\bibfnamefont{T.~R.} \bibnamefont{Kirkpatrick}},
  \bibnamefont{and} \bibinfo{author}{\bibfnamefont{T.}~\bibnamefont{Vojta}},
  \bibinfo{journal}{Phys. Rev. B} \textbf{\bibinfo{volume}{55}},
  \bibinfo{pages}{9452} (\bibinfo{year}{1997}).

\bibitem[{\citenamefont{Chitov and Millis}(2001)}]{Chitov_Millis_2001}
\bibinfo{author}{\bibfnamefont{G.~Y.} \bibnamefont{Chitov}} \bibnamefont{and}
  \bibinfo{author}{\bibfnamefont{A.~J.} \bibnamefont{Millis}},
  \bibinfo{journal}{Phys. Rev. B} \textbf{\bibinfo{volume}{64}},
  \bibinfo{pages}{054414} (\bibinfo{year}{2001}).

\bibitem[{\citenamefont{Chubukov and Maslov}(2003)}]{Chubukov_Maslov_2003}
\bibinfo{author}{\bibfnamefont{A.}~\bibnamefont{Chubukov}} \bibnamefont{and}
  \bibinfo{author}{\bibfnamefont{D.}~\bibnamefont{Maslov}},
  \bibinfo{journal}{Phys. Rev. B} \textbf{\bibinfo{volume}{68}},
  \bibinfo{pages}{155113} (\bibinfo{year}{2003}).

\bibitem[{\citenamefont{S{\'o}lyom}(1979)}]{Solyom_1979}
\bibinfo{author}{\bibfnamefont{J.}~\bibnamefont{S{\'o}lyom}},
  \bibinfo{journal}{Adv. Phys.} \textbf{\bibinfo{volume}{28}},
  \bibinfo{pages}{201} (\bibinfo{year}{1979}).

\bibitem[{\citenamefont{Lee and Ramakrishnan}(1985)}]{Lee_Ramakrishnan_1985}
\bibinfo{author}{\bibfnamefont{P.~A.} \bibnamefont{Lee}} \bibnamefont{and}
  \bibinfo{author}{\bibfnamefont{T.~V.} \bibnamefont{Ramakrishnan}},
  \bibinfo{journal}{Rev. Mod. Phys.} \textbf{\bibinfo{volume}{57}},
  \bibinfo{pages}{287} (\bibinfo{year}{1985}).

\bibitem[{\citenamefont{Belitz and
  Kirkpatrick}(1994)}]{Belitz_Kirkpatrick_1994}
\bibinfo{author}{\bibfnamefont{D.}~\bibnamefont{Belitz}} \bibnamefont{and}
  \bibinfo{author}{\bibfnamefont{T.~R.} \bibnamefont{Kirkpatrick}},
  \bibinfo{journal}{Rev. Mod. Phys.} \textbf{\bibinfo{volume}{66}},
  \bibinfo{pages}{261} (\bibinfo{year}{1994}).

\bibitem[{\citenamefont{Zinn-Justin}(1996)}]{Zinn-Justin_1996}
\bibinfo{author}{\bibfnamefont{J.}~\bibnamefont{Zinn-Justin}},
  \emph{\bibinfo{title}{Quantum Field Theory and Critical Phenomena}}
  (\bibinfo{publisher}{Oxford University Press, Oxford}, \bibinfo{year}{1996}).

\bibitem[{\citenamefont{Finkelstein}(1983)}]{Finkelstein_1983}
\bibinfo{author}{\bibfnamefont{A.~M.} \bibnamefont{Finkelstein}},
  \bibinfo{journal}{Zh. Eksp. Teor. Fiz.} \textbf{\bibinfo{volume}{84}},
  \bibinfo{pages}{168} (\bibinfo{year}{1983}), \bibinfo{note}{[Sov. Phys. JETP
  {\bf 57}, 97 (1983)]}.

\bibitem[{DOS()}]{DOS_footnote}
\bibinfo{note}{A DOS that vanishes on the Fermi surface and only there can be
  caused by competing mechanisms for establishing long-range order, none of
  which are successful because the system is at a lower critical dimension.
  This is often thought to be the reason for the vanishing DOS in a LL.
  However, this is not the only possibility. For instance, at the Anderson-Mott
  transition in $d>2$ there are no such competing mechanisms, although in $d=2$
  there are.}

\bibitem[{\citenamefont{Maleyev and Toperverg}(1975)}]{Maleyev_Toperverg_1975}
\bibinfo{author}{\bibfnamefont{S.~V.} \bibnamefont{Maleyev}} \bibnamefont{and}
  \bibinfo{author}{\bibfnamefont{B.~P.} \bibnamefont{Toperverg}},
  \bibinfo{journal}{Zh. Eksp. Teor. Fiz.} \textbf{\bibinfo{volume}{69}},
  \bibinfo{pages}{1440} (\bibinfo{year}{1975}), \bibinfo{note}{[Sov. Phys. JETP
  {\bf 42}, 734 (1976)]}.

\bibitem[{\citenamefont{Sch{\"a}fer and Wegner}(1980)}]{Schaefer_Wegner_1980}
\bibinfo{author}{\bibfnamefont{L.}~\bibnamefont{Sch{\"a}fer}} \bibnamefont{and}
  \bibinfo{author}{\bibfnamefont{F.}~\bibnamefont{Wegner}},
  \bibinfo{journal}{Z. Phys. B} \textbf{\bibinfo{volume}{38}},
  \bibinfo{pages}{113} (\bibinfo{year}{1980}).

\bibitem[{\citenamefont{Belitz and
  Kirkpatrick}(1997)}]{Belitz_Kirkpatrick_1997}
\bibinfo{author}{\bibfnamefont{D.}~\bibnamefont{Belitz}} \bibnamefont{and}
  \bibinfo{author}{\bibfnamefont{T.~R.} \bibnamefont{Kirkpatrick}},
  \bibinfo{journal}{Phys. Rev. B} \textbf{\bibinfo{volume}{56}},
  \bibinfo{pages}{6513} (\bibinfo{year}{1997}).

\bibitem[{mul()}]{multiple_z_footnote}
\bibinfo{note}{The thermodynamics are dominated by the critical modes, so we
  expect the relevant dynamical exponent to be $z$. However, in general all
  scaling functions depend on both $z$ and $z_{\text{c}}$, and scaling
  considerations alone cannot determine which one enters in any given context.
  The answer depends on the relative values of the various dynamical exponents,
  the coupling strengths between modes, and the presence of dangerous
  irrelevant variables.}

\bibitem[{\citenamefont{Sachdev}(1999)}]{Sachdev_1999}
\bibinfo{author}{\bibfnamefont{S.}~\bibnamefont{Sachdev}},
  \emph{\bibinfo{title}{Quantum Phase Transitions}}
  (\bibinfo{publisher}{Cambridge University Press, Cambridge},
  \bibinfo{year}{1999}).

\bibitem[{\citenamefont{Belitz et~al.}(2002)\citenamefont{Belitz, Kirkpatrick,
  and Vojta}}]{Belitz_Kirkpatrick_Vojta_2002}
\bibinfo{author}{\bibfnamefont{D.}~\bibnamefont{Belitz}},
  \bibinfo{author}{\bibfnamefont{T.~R.} \bibnamefont{Kirkpatrick}},
  \bibnamefont{and} \bibinfo{author}{\bibfnamefont{T.}~\bibnamefont{Vojta}},
  \bibinfo{journal}{Phys. Rev. B} \textbf{\bibinfo{volume}{65}},
  \bibinfo{pages}{165112} (\bibinfo{year}{2002}).

\bibitem[{z_c()}]{z_c_footnote}
\bibinfo{note}{Here we explicitly show the two time scales in the scaling
  function. Equation (\ref{eq:8b}) follows under certain assumptions, see Ref.\
  \onlinecite{multiple_z_footnote}, $z < 1$ being one of them. From scaling
  alone we cannot rule out $\rho(t=0,T) \propto T^{1/z}$.}

\bibitem[{\citenamefont{Altshuler and Aronov}(1979)}]{Altshuler_Aronov_1979}
\bibinfo{author}{\bibfnamefont{B.~L.} \bibnamefont{Altshuler}}
  \bibnamefont{and} \bibinfo{author}{\bibfnamefont{A.~G.}
  \bibnamefont{Aronov}}, \bibinfo{journal}{Solid State Commun.}
  \textbf{\bibinfo{volume}{30}}, \bibinfo{pages}{115} (\bibinfo{year}{1979}).

\bibitem[{\citenamefont{Belitz and
  Kirkpatrick}(1995)}]{Belitz_Kirkpatrick_1995}
\bibinfo{author}{\bibfnamefont{D.}~\bibnamefont{Belitz}} \bibnamefont{and}
  \bibinfo{author}{\bibfnamefont{T.~R.} \bibnamefont{Kirkpatrick}},
  \bibinfo{journal}{Z. Phys. B} \textbf{\bibinfo{volume}{98}},
  \bibinfo{pages}{513} (\bibinfo{year}{1995}).

\bibitem[{us_()}]{us_tbp}
\bibinfo{note}{D. Belitz and T.~R. Kirkpatrick, unpublished results.}

\bibitem[{\citenamefont{Br{\'e}zin and Wallace}(1973)}]{Brezin_Wallace_1973}
\bibinfo{author}{\bibfnamefont{E.}~\bibnamefont{Br{\'e}zin}} \bibnamefont{and}
  \bibinfo{author}{\bibfnamefont{D.~J.} \bibnamefont{Wallace}},
  \bibinfo{journal}{Phys. Rev. B} \textbf{\bibinfo{volume}{7}},
  \bibinfo{pages}{1967} (\bibinfo{year}{1973}).

\bibitem[{\citenamefont{Wegner}(1979)}]{Wegner_1979}
\bibinfo{author}{\bibfnamefont{F.}~\bibnamefont{Wegner}}, \bibinfo{journal}{Z.
  Phys. B} \textbf{\bibinfo{volume}{35}}, \bibinfo{pages}{207}
  (\bibinfo{year}{1979}).

\bibitem[{z_f()}]{z_footnote}
\bibinfo{note}{An argument given by J. Cardy, {\it Scaling and Renormalization
  in Statistical Physics} (Cambridge University Press, Cambridge, 1996), Sec.
  6.5, adapted to the current problem, suggests $\beta = O(1/\sqrt{\epsilon})$,
  or $z = 1 - O(\sqrt{\epsilon})$. An explicit 2-loop calculation is needed to
  distinguish between these scaling scenarios. In either case $z < z_{\text{c}}
  = 1$, which is one of the assumptions that yield Eq.\ (\ref{eq:8b}).}

\bibitem[{\citenamefont{Parisi}(1980)}]{Parisi_1980}
\bibinfo{author}{\bibfnamefont{G.}~\bibnamefont{Parisi}}, \bibinfo{journal}{J.
  Stat. Phys.} \textbf{\bibinfo{volume}{23}}, \bibinfo{pages}{49}
  (\bibinfo{year}{1980}).

\end{thebibliography}

\end{document}